\documentstyle[12pt]{article}

\begin{document}

\def\BbbZ{Z\!\!\!Z}

\hoffset = -1truecm
\voffset = -2truecm





\title {\bf A Quantum Gauge Group Approach\\
to the 2D SU(n) WZNW Model
}

\author{
{\bf Paolo Furlan}\\
\normalsize Dipartimento di Fisica Teorica dell'Universit\`a di Trieste,
Trieste,
{\bf Italy}\\
{\normalsize and}\\
\normalsize Istituto Nazionale di Fisica Nucleare (INFN), Sezione di
Trieste, Trieste,
{\bf Italy,} \\
{\bf Ludmil K. Hadjiivanov\thanks{On leave of absence from:
Division of Theoretical Physics,
Institute for Nuclear Research and Nuclear Energy, Bulgarian Academy of
Sciences, Tsarigradsko Chaussee 72, 1784 Sofia, Bulgaria.}}\\
\normalsize International Centre for Theoretical Physics, 34100 Trieste,
{\bf Italy}\\
{\normalsize and}\\
\normalsize Istituto Nazionale di Fisica Nucleare (INFN), Sezione di
Trieste, Trieste,
{\bf Italy}\\
{\normalsize and}\\
{\bf Ivan T. Todorov}\\
\normalsize Division of Theoretical Physics, Institute for
Nuclear Research and Nuclear Energy,\\
\normalsize Bulgarian Academy of Sciences, Sofia,
{\bf Bulgaria} \\
{\normalsize and}\\
\normalsize Scuola Internazionale Superiore di Studi Avanzati (SISSA/ISAS),\\
\normalsize 34014 Trieste, {\bf Italy}}

\date{21st October 1996; Revised 
6th March 1997}


\maketitle


\begin{abstract}
The canonical quantization of the WZNW model
provides a complete set of exchange relations
in the enlarged chiral state spaces that include the Gauss components
$M_\pm$, ${\bar M}_\pm$  of the monodromy matrices $M$, $\bar M$.
Regarded as new dynamical variables, the elements of $M$ and $\bar M$
cannot be identified -- they satisfy different exchange relations.
Accordingly, the two dimensional theory expressed in terms of
the left and right movers' fields does not
automatically respect monodromy invariance.

Continuing our recent
analysis of the problem by gauge theory methods we conclude that physical
states $\Phi$ (on which the field $u(x-t){\bar u}(x+t) \in SU(n)$ acts as
a single valued operator) are invariant under the (permuted) coproduct of
the left and right $U_q(sl(n))\,.$
They satisfy additional constraints fully described for $n=2$; then $(i)\,
\Phi$ is invariant under a second $U_q(sl(2))$, which commutes with the
coproduct, and $(ii) (A^-)^{h-1}(A^+)^l\,\Phi = 0\ ,\
l=0,1,\ldots ,h-2$, where $A^{\pm}$ are quantum
group invariant creation and annihilation operators 
such that $(A^{\pm})^h = 0$ ($h=k+2$ being the height).

\end{abstract}

\newpage


\section{Introduction and Summary}
\noindent
The general group valued periodic solution $g(t,x+2\pi )=g(t,x)$ of the
Wess--Zumino--Novikov--Witten (WZNW) equations of motion factorizes into
a product of   "right and left movers",
\begin{equation}
g(t,x)=u(x-t){\bar u}(x+t)\,\,,\,\,\, g, u, {\bar u}\in G
\label{1}
\end{equation}
satisfying a {\it {twisted periodicity}} condition [1,2]:
\begin{equation}
u(x+2\pi )=u(x)M\,,\,\,\,{\bar u}(x+2\pi )={\bar M}^{-1}{\bar u}(x)\,.
\label{2}
\end{equation}

The symplectic form of the $2D$ theory can be presented as a sum of two
decoupled, closed, chiral $2$-forms at the price of considering the
monodromy matrices $M$ and $\bar M$ as independent of each other
additional dynamical variables [2,3]. Restricting  attention to the
case $G=SU(n)\,$ one then derives [3] quadratic
Poisson bracket relations involving the components $M_\pm$ of the Gauss
decomposition of $M\,$:
\begin{eqnarray*}
M=M_+M_-^{-1}\,,\quad M_\pm^{\pm 1}=
N_\pm D\,,
\end{eqnarray*}
\begin{equation}
D=(d_i\delta_{ij})\ ,\,
N_+= \left(\matrix{ 1 &f_1 &f_{12} &\ldots\cr
0 &1 &f_2 &\ldots\cr
0 &0 &1 &\ldots\cr
\ldots &\ldots &\ldots &\ldots
\cr}\right),\,
N_-= \left(\matrix{ 1 &0 &0 &\ldots\cr
e_1 &1 &0 &\ldots\cr
e_{21} &e_2 &1 &\ldots\cr
\ldots &\ldots &\ldots &\ldots\cr}\right)\,, \label{3}
\end{equation}
where the common diagonal matrix $D$ has unit determinant,
\begin{equation}
d_1d_2\ldots d_n=1
\end{equation}
(and the same for the bar sector). The form of the Poisson brackets implies
upon quantization quadratic exchange relations [2,3,4]:
\begin{eqnarray*}
\stackrel{2}{u}(x_2)\stackrel{1}{u}(x_1)\,=\,\stackrel{1}{u}(x_1)\stackrel{2}{u}
(x_2) R(x_{12})\,,\,\,\,
\stackrel{2}{\bar u}(x_2)\stackrel{1}{\bar u}(x_1)\,=\,
\bar R(x_{21})\stackrel{1}{\bar u}(x_1)\stackrel{2}{\bar u}(x_2)\,,
\label{5}
\end{eqnarray*}
\begin{equation}
x_{ij}=x_i-x_j\ne 0\,,\quad
R(x)=R^-\ \theta (x)+R^+\ \theta (-x)\,,\quad
\bar R(-x)=\left(R(x)\right)^{-1}\,.
\end{equation}
Here $\theta$ is the step function with periodically continued
derivative, $2\pi{\theta}'(x)=\delta (x) =\sum_n e^{inx}\,$, and $R^\pm$
are the constant solutions of the quantum Yang--Baxter equation
corresponding to $SL_q(n)$ [6]:
\begin{eqnarray*}
R^\varepsilon_{12}\ R^\pm_{13}\ R^\pm_{23}=R^\pm_{23}\ R^\pm_{13}\
R^\varepsilon_{12}\quad\varepsilon =+\ , - \,;
\end{eqnarray*}
\begin{equation}
( R^\pm )^{\alpha\beta}_{{\alpha}'{\beta}'}=q^{\pm {1\over
n}}\left(q^{\mp{\delta}_{\alpha
\beta}}{\delta}^{\alpha}_{{\alpha}'}{\delta}^{\beta}_{{\beta}'}\pm
(\bar q -q){\theta}_{{\beta}^>_< \alpha} {\delta}^{\alpha}_{{\beta}'}
{\delta}^{\beta}_{{\alpha}'}\right)\,.
\end{equation}
The monodromy matrices $M_\pm$ and ${\bar M}_\pm^{-1}$ satisfy
{\it {identical exchange relations}} (so that $M\ne \bar M$ and
hence the monodromy invariance of $g$ given by Eq.~(1) is {\it{not}}
automatic): \begin{eqnarray}
R^\varepsilon
\stackrel{1}M_\pm\stackrel{2}M_\pm\,=\,\stackrel{2}M_\pm\stackrel{1}M_\pm
R^\varepsilon \,&,&\quad
R^\varepsilon
\stackrel{1}{\bar M}_\pm\!\!\! ^{-1}\stackrel{2}{\bar M}_\pm\!\!\! ^{-1}\,=\,
\stackrel{2}{\bar M}_\pm\!\!\! ^{-1}\stackrel{1}{\bar M}_\pm\!\!\! ^{-1}
R^\varepsilon \,,
\nonumber \\
R^+\stackrel{1}M_+\stackrel{2}M_-\,=\,\stackrel{2}M_-\stackrel{1}M_+R^+\,&,&
\quad
R^+\stackrel{1}{\bar M}_+\!\!\! ^{-1}\stackrel{2}{\bar M}_-\!\!\! ^{-1}\,=\,
\stackrel{2}{\bar M}_-\!\!\! ^{-1}\stackrel{1}{\bar M}_+\!\!\! ^{-1}R^+\,,
\end{eqnarray}
while their relations with $u$ and $\bar u$ differ:
\begin{equation}
\stackrel{1}{M}_\pm\stackrel{2}{u}(x)\,=\,
\stackrel{2}{u}(x)R^\pm\stackrel{1}{M}_\pm\,,\quad
\stackrel{1}{\bar M}_\pm\!\!\!^{-1}R^\pm\stackrel{2}{\bar u}(x)\,=\,
\stackrel{2}{\bar u}(x)\stackrel{1}{\bar M}_\pm\!\!\!^{-1}
\end{equation}
(one also has $[\stackrel{1}{M}_\varepsilon\,,\stackrel{2}{\bar M}_
{{\varepsilon}'}]\,=\,[\stackrel{1}{u}(x)\,,\stackrel{2}{\bar u}(y)]\,=\,
[\stackrel{1}{M}_\varepsilon\,,\stackrel{2}{\bar u}(x)]
\,=\,[\stackrel{1}{u}(x)\,,\stackrel{2}{\bar M}_\varepsilon]\,=\,0\,$).

In [5] we introduced  a realization for the chiral vertex operators
$u(x)$ in the case of $SU(2)$ in terms of
$U_q^{(2)}:=U_q(sl(2))$ oscillators
$a^\pm_\beta$ and $\bar a^\alpha_\pm\,$:
\begin{equation}
u(x)^\alpha_\beta =u^\alpha_+(x,N)a^+_\beta +a^-_\beta
u^\alpha_-(x,N)\,,\quad
{\bar u}(x)^\alpha_\beta ={\bar u}^+_\beta(x,\bar N){\bar a}^\alpha_+
+{\bar a}^\alpha_-{\bar u}^-_\beta (x,\bar N)\,.
\end{equation}
Here $N\,,\bar N$ are modified number operators (defined modulo $2h$
and shifted by $1\,$):
\begin{eqnarray}
q^Na^\pm_\alpha =a^\pm_\alpha q^{N\pm 1}\,,\quad
q^{\bar N}{\bar a}^\alpha_\pm
={\bar a}^\alpha_\pm q^{{\bar N}\pm 1}\,, \nonumber \\
\left[ N\,,{\bar a}^{\alpha}_{\varepsilon}\right]=0=
\left[{\bar N}\,,a^{\varepsilon}_{\beta}\right]\,,\,\,
q^N|0>=q|0>=q^{\bar N}|0>\,.
\end{eqnarray}
The $U_q^{(2)}$ properties of the oscillators are
coded (as discussed in Sec. $2$ ) in the exchange relations
\begin{equation}
(M_\pm )^\alpha_\beta a^\varepsilon_\gamma =
a^\varepsilon_\tau (R^\pm )^{\alpha\tau}_{\sigma\gamma}(M_\pm
)^\sigma_\beta\,,
\quad ({\bar M}^{-1}_\pm )^\alpha_\sigma (R^\pm
)^{\sigma\gamma}_{\beta\tau}{\bar a}^\tau_\varepsilon =
{\bar a}^\gamma_\varepsilon ({\bar M}^{-1}_\pm )^\alpha_\beta\,,
\quad\varepsilon =+\ , - \,.
\end{equation}
Eqs. (5), on the other hand, are implemented by 
the $U_q^{(2)}$-Bose commutation
relations
\begin{eqnarray}
a^\varepsilon_\alpha a^\varepsilon_\beta = q^{\pm{1\over 2}}\
a^\varepsilon_\rho a^\varepsilon_\sigma\ (\check R^{\pm} )^{\rho\sigma}
_{\alpha\beta}\,\,&,&\quad
a^-_\alpha a^+_\beta = q^{\pm{1\over 2}} a^+_\rho a^-_\sigma\ (\check
R^{\pm} )^{\rho\sigma}_{\alpha\beta} +q^{\mp (N-1)}{\cal
E}_{\alpha\beta}\,, \nonumber \\
\bar a^\alpha_\varepsilon\ \bar a^\beta_\varepsilon = q^{\pm{1\over 2}}\
(\check R^{\pm} )^{\alpha\beta}_{\rho\sigma}\ \bar a^\rho_\varepsilon\ \bar
a^\sigma_\varepsilon \,\,&,&\quad
\bar a^\alpha_-\ \bar a^\beta_+ = q^{\pm{1\over 2}}\
(\check R^{\pm} )^{\alpha\beta}_{\rho\sigma}\ \bar a^\rho_+\ \bar
a^\sigma_--q^{\mp (\bar N-1)} \ {\cal E}^{\alpha\beta}
\label{12}
\end{eqnarray}
where $({\cal E}_{\alpha\beta})$ and $({\cal E}^{\alpha\beta})$ are the
$SL_q(2)$ invariant tensor and its inverse,
\begin{equation}
\left({\cal E}_{\alpha\beta}\right) =\left(\matrix{ 0 &-q^{1/2}\cr
\bar q^{1/2} &0\cr}\right) = \left( -{\cal E}^{\alpha\beta}\right) \,,\quad
-{\cal E}^{\alpha\beta}{\cal E}_{\alpha\beta}=[2]:=q+\bar q =2cos{\pi\over
h}
\end{equation}
($\bar q q=1\,, q^h=-1$).~${\check R}^+$ is the braid operator
and ${\check R}^-$ its inverse, 
\begin{equation}
{\check R}^+ =R^+P = \check R\,,\quad 
{\check R}^-=R^-P={\check R}^{-1}
\,;\quad
P^{\alpha\beta}_{{\alpha}'{\beta}'}=
{\delta}^\alpha_{{\beta}'}{\delta}^\beta_{{\alpha}'}\,.
\end{equation}
Using further the expressions for the (two sets of) quantum universal
enveloping algebra generators in terms of the $U_q^{(2)}$ oscillators (that
can be verified in the Fock space representation of the basic CR),
\begin{eqnarray}
E=-{\bar q}^{1\over 2}a^+_1a^-_1\,,\,\,
F=a^+_2a^-_2q^{{3\over 2}-H}\,&,&\,\,
\bar E=-{\bar a}^2_+{\bar a}^2_-q^{{1\over 2}+\bar H}\,,\,\,
\bar F=q^{1\over 2}{\bar a}^1_+{\bar a}^1_-\,,\\
(q^{3\over 2}-{\bar q}^{1\over 2})a^{\pm}_1a^{\mp}_2=
q^H - q^{\mp N+1}\,&,&\,\,
(q^{1\over 2}-{\bar q}^{3\over 2})a^{\pm}_2a^{\mp}_1=
q^H - q^{\pm N-1}\,,\\
(q^{3\over 2}-{\bar q}^{1\over 2}){\bar a}^1_{\pm}{\bar a}^2_{\mp}=
{\bar q}^{\bar H} - q^{\mp{\bar N}+1}\,&,&\,\,
(q^{1\over 2}-{\bar q}^{3\over 2}){\bar a}^2_{\pm}{\bar a}^1_{\mp}=
{\bar q}^{\bar H} - q^{\pm{\bar N}-1}\,,
\end{eqnarray}
( where $[x]:={{q^x-q^{-x}}\over {q-q^{-1}}}\,$) and
\begin{equation}
q^Ha^{\varepsilon}_1=a^{\varepsilon}_1q^{H+1}\,,\,\,
q^Ha^{\varepsilon}_2=a^{\varepsilon}_2q^{H-1}\,,\,\,
q^{\bar H}{\bar a}_{\varepsilon}^1=
{\bar a}_{\varepsilon}^1q^{{\bar H}-1}\,,\,\,
q^{\bar H}{\bar a}_{\varepsilon}^2=
{\bar a}_{\varepsilon}^2q^{{\bar H}+1}\,,
\end{equation}
we have expressed in [5] (note change of conventions for the bar sector)
the two monodromy matrices in terms of the corresponding
$U_q^{(2)}$ generators thus revealing the dynamical role of the quantum
groups in a WZNW theory. It has been demonstrated that $U_q^{(2)}$ plays the
role of a gauge symmetry in the "physical subspace" ${\cal H}'$ whose
quantum oscillator part is generated by polynomials of the $U_q^{(2)}$
invariant creation operator $A^+=a^+_\alpha {\bar a}^{\alpha}_+$ on the
vacuum. Monodromy invariance is satisfied in a weak sense:
\begin{equation}
\{u(x)M{\bar M}^{-1}{\bar u}(\bar x)-u(x){\bar u}({\bar x})\}{\cal H}'
= 0\,.
\end{equation}
The question of writing down a full set of constraints which single out
the subspace ${\cal H}'$ has been left open in [5].

The objective of the present paper is twofold. First, we demonstrate (in
Sec. $2$) that the monodromy matrices for an arbitrary $SU(n)$ WZNW model are
expressed in terms of the $U_q^{(n)}:=U_q(sl(n))$ generators. This is derived
as a direct consequence of the exchange relations (7) without the
intermediation of quantum oscillators. We show that the matrix
operators $L_+$ and $L_-$ where
\begin{equation}
L_+={\bar M}^{-1}_+M_+\,,\quad L_-={\bar M}_-^{-1}M_-
\quad ({\rm satisfying}\quad [\,L_\pm\,,\,u(x){\bar u}(\bar x )\,]=0\,)
\end{equation}
give rise to the (permuted) coproduct of the chiral $U_q^{(n)}$ generators.
Secondly, for $n=2$ we show (in Sec. $4$) that $U_q^{(2)}$ invariance is not
enough
to single out the physical subspace ${\cal H}'\,$. In fact, there is a
second $U_q^{(2)}$ algebra (commuting with $L_\pm$ and with the observables)
generated by $a^+_\alpha {\bar a}^\alpha_-\,,\, a^-_\alpha {\bar a}^\alpha_+$
and $q^{\pm (N-{\bar N})}$ satisfying
\begin{equation}
[a^-_\alpha {\bar a}^\alpha_+\,, a^+_\alpha {\bar a}^\alpha_- ]=
[N-{\bar N}]\,,\quad
q^{N-{\bar N}} a^{\pm}_\alpha {\bar a}^\alpha_{\mp}=
a^{\pm}_\alpha {\bar a}^\alpha_{\mp}q^{N-{\bar N}\pm 2}\,.
\end{equation}
The subspace ${\cal H}'$ is obtained from the tensor product space
${\cal H}\otimes{\bar{\cal H}}$
of the state spaces of left and right movers in
terms of the conditions
\begin{eqnarray}
\{ (L_\pm)^\alpha_\beta - {\delta}^\alpha_\beta\}{\cal H}' =0\,,\quad
a^\mp {\bar a}_\pm{\cal H}' =0= \{q^{N-{\bar N}} -1\}{\cal H}'\,,
\nonumber \\
(a^- {\bar a}_-)^{h-1} ({a^+ {\bar a}_+} )^l\ {\cal H}' =0\ ,
\ l=0,1,\ldots ,h-2\ ;
\,\,\,\,\, a^{\varepsilon} {\bar a}_{\varepsilon'} \equiv
a^{\varepsilon}_{\alpha} {\bar a}^{\alpha}_{\varepsilon'} \, .
\end{eqnarray}

It is only in ${\cal H}'$ that the "physical inner product"
$\langle\,.\,\vert\,.\,\rangle$ (introduced in Sec. $3$) is positive
semidefinite. The appearance of the two commuting $U_q^{(2)}\,$ algebras
should be viewed as a manifestation of the regular representation
of $U_q^{(2)}\,$ (which is finite dimensional for $q$ a root of unity).
Hence, they admit a generalization to $U_q^{(n)}$ for arbitrary $n\,$.

\vspace{1pt}
\section{\boldmath{$U_q(sl(n))$ Symmetry in the $SU(n)$ WZNW Model}}
The quantized $u$ (and $g$) cannot be treated as group elements. Instead
of setting $u\in SU(n)$ we can just assert that the operator product
expansion of $u^*(x)u(y)$ only involves fields of the family (or, rather,
the Verma module) of the unit operator. The relation
$e^{-2\pi iL_0}u(x)e^{2\pi iL_0}=u(x)M$ implies
$M^{\alpha}_{\beta}\vert 0\rangle
={\bar q}^{{n^2-1}\over n}{\delta}^{\alpha}_{\beta}\vert 0\rangle\,$
 (cf. [4,5]); hence, the ("quantum") determinant of the monodromy differs
from $1$. As a result, the representation (3) is only valid if we
substitute the first equation by $q^{{n^2-1}\over n}M=M_+M^{-1}_-\,$.

\vspace{12pt}
\noindent
{\bf Proposition~2.1:} {\it The CR (7) are satisfied for $d_i, f_i, f_{ij},
e_i, e_{ji}$ expressed in terms of the Chevalley--Cartan generators of
$U_q^{(n)}\,$:}
\begin{equation}
d_i=q^{{\Lambda}^{\vee}_{i-1}-{\Lambda}^{\vee}_i}, i=1,2,\ldots ,n \,\,
({\Lambda}^{\vee}_0=0={\Lambda}^{\vee}_n),\,\,
e_i=(\bar q-q)E_i\,,\, f_i=(\bar q-q )F_i\,\,,
\end{equation}
\begin{eqnarray*}
(\bar q-q)e_{j+1~j}=e_je_{j+1}-qe_{j+1}e_j\,,\,
(\bar q-q)e_{j+1~i}=e_{ji}e_{j+1}-qe_{j+1}e_{ij}\,\,,
\end{eqnarray*}
\begin{equation}
(\bar q-q)f_{i~i+1}=f_{i+1}f_i-qf_if_{i+1}\,,\,\,
(\bar q-q)f_{i~j+1}=f_{j+1}f_{ij}-qf_{ij}f_{j+1}\quad (i\le j)\,\,.
\end{equation}
{\it Here ${\Lambda}^{\vee}_j$ are the fundamental coweights of $su(n)$
characterized by $(H_i,{\Lambda}^{\vee}_j) = {\delta}_{ij}\,$:}
\begin{equation}
H_i=2{\Lambda}^{\vee}_i-{\Lambda}^{\vee}_{i-1}-{\Lambda}^{\vee}_{i+1}\,,\quad
q^{{\Lambda}^{\vee}_i}E_j=E_jq^{{\Lambda}^{\vee}_i+{\delta}_{ij}}\,,\quad
q^{{\Lambda}^{\vee}_i}F_j=F_jq^{{\Lambda}^{\vee}_i-{\delta}_{ij}}
\end{equation}
{\it ($\,\{H_i\}$ being the Cartan basis of simple coroots of $su(n)\,$),
and the raising and lowering operators $E_i\,,\,F_j$ obey the CR}
\begin{eqnarray}
[E_i\,,\,F_j]=[H_i]{\delta}_{ij}\,,\quad
[E_i\,,\,E_j]=0=[F_i\,,\,F_j]\,\,{\rm for}\,\, |i-j|\ge 2\,,
\nonumber \\
\left[ 2\right] E_iE_{i\pm 1}E_i=E_{i\pm 1}E_i^2+E_i^2E_{i\pm 1}\,,\quad
\left[ 2\right] F_iF_{i\pm 1}F_i=F_{i\pm 1}F_i^2+F_i^2F_{i\pm 1}\,.
\end{eqnarray}
{\it The same type of parametrization can be written for ${\bar
M}_+^{-1}$ and
${\bar M}_-\,$. The products $L_{\pm}$ (Eq. (20)) give rise to the
(permuted) coproduct of $U_q^{(n)}\,$:}
\begin{equation}
\Delta (E_i)=q^{H_i}\otimes E_i +E_i\otimes 1\,,\quad
\Delta (F_i)=F_i\otimes {\bar q}^{H_i}+1\otimes F_i\,,\quad
\Delta (q^{H_i})=q^{H_i}\otimes q^{H_i}
\end{equation}
{\it where the second tensor multipliers act on the bar sector.}

\vspace{12pt}
\noindent
{\bf Proof:} The exchange relations (7) can be rewritten in the form
\begin{equation}
[d_i\,,d_j]=0\,,\,\,
d_i\,e_j=q^{{\delta}_{ij+1}-{\delta}_{ij}}e_j\,d_i\,,\,\,
d_i\,f_j=q^{{\delta}_{ij}-{\delta}_{ij+1}}f_j\,d_i
\end{equation}
\begin{equation}
[e_i\,,f_j]=(q-\bar
q)(d_i^{-1}d_{i+1}-d_i\,d_{i+1}^{-1})\,
{\delta}_{ij}\,,\,\,\,
e_i\,e_{ji}=q\,e_{ji}\,e_i\,\,(\,{\rm for}\,\, j>i\,)\,\,{\rm etc.}
\end{equation}
where $e_{ji}$ etc. are defined in (24). Eq. (29) is derived from a
reflected version of the second equation (7):
\begin{equation}
\stackrel{1}{M}_+R^+\stackrel{2}{M}_-\!\!\!^{-1}=
\stackrel{2}{M}_-\!\!\!^{-1}R^+\stackrel{1}{M}_+\,,
\end{equation}
which implies, in particular,
\begin{eqnarray*}
[f_i\,d_{i+1}\,,e_{i+1}d_{i+1}]=0\,,\,\,\,
[e_i\,d_i\,,f_i\,d_{i+1}]=(q^2-1)(d^2_{i+1}-d^2_i\,)\,.
\end{eqnarray*}
The commutativity of $d_i$ is implemented by their realization (23)
in terms of the (commuting) fundamental coweights. The remaining Eqs.
(28) are equivalent to (25). The CR (26) are a realization of (29) .

\vspace{12pt}
\noindent
{\bf Remark:}
The term {\it {permuted}} for the coproduct (27) is justified by the
fact that it is obtained by a permutation from the standard coproduct
[6]; in fact, we can write
$\left({L_\pm}\right)^\alpha_\beta =
({{\stackrel{2}{\cal M}}_\pm} )^\alpha_\sigma
({{\stackrel{1}{\cal M}}_\pm} )^\sigma_\beta\,$ where
$\,({{\stackrel{2}{\cal M}}_\pm} )^\alpha_\sigma =
1\otimes ({\bar M}_\pm^{-1} )^\alpha_\sigma\,,\,
({{\stackrel{1}{\cal M}}_\pm} )^\sigma_\beta =
({M_\pm} )^\sigma_\beta \otimes 1\,.$
Since this is the only coproduct we will be dealing with, we do not use a
special sign (like ${\Delta}'\,$)
for it. Its choice is dictated by the fact that only $L_\pm$
given by Eq.~(20) ({\it {not}} $M_+{\bar M}_+^{-1}\,$ etc.) commute with
$u\bar u\,$.

\section{Hermitean Conjugation and Transposition. Physical Inner Product}
\noindent
The associative algebra $U_q^{(n)}$ admits for $q$ on the unit circle two
antiinvolutions (that leave the basic CR invariant):
an {\it{antilinear hermitean conjugation}} $X\to X^*\,$, and a
{\it{linear transposition}} $X\to {^tX}\,$ acting on the
Chevalley--Cartan generators as
\begin{equation}
(q^{H_i})^*={\bar q}^{H_i}\,,\quad E_i^*=F_i\,,\quad F_i^*=E_i
\quad (q^*=\bar q)\,,
\end{equation}
\begin{equation}
^t(q^{H_i})=q^{H_i}\,,\,\, ^tE_i=F_iq^{H_i-1}\,,\,\,
^tF_i={\bar q}^{H_i-1}E_i\,\,
\Leftrightarrow \,\, ^tM_+=M^{-1}_-\,,
\end{equation}
respectively. The star becomes a bialgebra antihomomorphism if we set
$(X\otimes Y)^*=Y^*\otimes X^*$ while the transposition preserves the
coproduct provided $^t(X\otimes Y)={^tX}\otimes{^tY}\,$. The
transposition is
determined by its properties up to a (cyclic) inner automorphism,
\begin{equation}
X\to q^{m{\rho}^{\vee}}X{\bar q}^{m{\rho}^{\vee}}\,,\,\,\,{\rho}^{\vee}
=\sum_i {\Lambda}^{\vee}_i\,\,\,
(\,\Rightarrow\,\,E_i\to q^mE_i\,,\,\,F_i\to {\bar q}^mF_i\,)\,.
\end{equation}
Our choice fits the last Eq.~(32).

The conjugation (31) gives rise to a $U_q^{(n)}$ invariant {\it{hermitean
form}} $(\,.\,\vert\,.\,)$ in each (finite dimensional)
irreducible highest
weight module ${\cal V}_\Lambda$ (with highest weight vector ${\vert\Lambda
\rangle\,}$) of $U_q^{(n)}\,$. It is uniquely determined by the conditions
\begin{equation}
(\,\Lambda \,\vert\Lambda\,)=1\,,\,\,(\,\Phi \vert X\Psi\,)=(\,X^*\Phi
\vert \Psi \,)\,\,
{\rm for}\,\,\, \Phi\,,\Psi\,\in {\cal V}_\Lambda\,, \,\,X\in U_q^{(n)}\,.
\end{equation}
It is linear in $\Psi\,$, antilinear in $\Phi$ and satisfies the hermiticity
condition $(\,\Phi \vert\Psi\,) =
{\overline {(\,\Psi \vert\Phi\,)}\,}$.
For $q=e^{i{\pi\over h}}$ the resulting scalar product is
positive definite for
\begin{equation}
(\theta\,,\Lambda) = {\lambda}_1+{\lambda}_2+\ldots
+{\lambda}_{n-1}
\le h-n+1\quad (\,\Lambda =\sum_{i=1}^{n-1} {\lambda}_i{\Lambda}_i ).
\end{equation}
Similarly, the transposition (32) gives rise to a non-degenerate symmetric
(complex valued) {\it{bilinear form}}
$\langle\,.\,\vert\,.\,\rangle$ on each ${\cal V}_\Lambda\,$, defined by
\begin{equation}
\langle\,\Lambda \,\vert\Lambda\,\rangle
\,=1\,,\,\,\langle\,\Phi \vert X\Psi\,\rangle\,=\,
\langle\,{^tX}\Phi\vert\Psi \,\rangle\,,\,\,\langle\,\Phi
\vert\Psi\,\rangle = \langle\,\Psi \vert\Phi\,\rangle\,.
\end{equation}
It can be shown
that for unitarizable representations (i.e., for highest weights
satisfying (35)) one has
\begin{equation}
|\langle\,\Phi\,\vert\Phi\,\rangle|\,\le\,
(\,\Phi\,\vert\,\Phi\,) \quad
{\rm{for~ all}}\,\,\Phi\in {\cal V}_\Lambda\,.
\end{equation}
It is the bilinear form (36) which plays the role of a {\it {physical
inner product}}. It is positive semidefinite in a suitably defined subspace
${\cal H}'$ of the extended $2D$ state space of the WZNW model.

We conjecture that each of the chiral state spaces
$\cal H$ and ${\bar {\cal H}}$
can be presented in the form of a finite sum of tensor products, e.g.,
\begin{equation}
{\cal H}={\oplus}_\Lambda {\cal H}_\Lambda\otimes {\cal V}_\Lambda
\quad(\,\,(\theta\,,\,\Lambda )\, \le h-n+1\,)\,,
\end{equation}
every unitarizable $\Lambda$ appearing with a finite multiplicity. We will
prove this conjecture and
display the finite dimensional {\it {internal chiral space}}
${\cal F} =
{\oplus}_{\Lambda} {\cal V}_\Lambda$
in the simplest case $n=2$.

We introduce an orthogonal basis $\{\,\vert mn\rangle\,\}$ in
${\cal F}= {\cal F}(U_q^{(2)})$
with the following range of the $U_q^{(2)}$ invariant sum of nonnegative
integers $m$ and $n\,$:
\begin{equation}
1\le N:=m+n+1\le 2h-1
\end{equation}
while their difference $\lambda :=m-n$ runs in the interval
\begin{equation}
|\lambda | \le{\rm{min}}\,(N-1\,,\,2h-N-1\,)\,.
\end{equation}
(each representation of highest weight $2I\le h-2$ is thus encountered twice.)
The $U_q^{(2)}$ generators are assumed to act in this basis
according to
\begin{equation}
E\vert mn\rangle
=[n]\vert m+1~n-1\rangle\,,\,\,
F\vert mn\rangle =[m]\vert m-1~n+1\rangle\,,\,\,
(q^H-q^\lambda )\vert mn\rangle=0\,. \end{equation}
(The vectors in the right hand side for which condition (40) is violated
are defined to be zero.)
It follows that $(EF-[m][n+1])\vert mn\rangle =0= (FE-[m+1][n])\vert
mn\rangle\,,$ in accord with Eq. (16) and Eq. (43) below.

This $h^2$-dimensional space is invariant under the
Weyl inversion
\begin{equation}
N \to 2h-N\,,\,\,\lambda \to -\lambda\quad(E \leftrightarrow F\,,\,\,H
\to -H\,).
\end{equation}
It can be realized in terms of the
$U_q^{(2)}$ Bose creation operators of Sec. 1 setting
\begin{equation}
|mn>=(a^+_1)^m (a^+_2)^n \vert 0\rangle\,,\,\,\,(a^\varepsilon_\alpha )^h =
0\,. \end{equation}
For the vectors of this basis the bilinear form (36) is proportional to its
hermitean counterpart (34):
\begin{equation}
\langle m'n' \vert mn\rangle={\delta}_{mm'}{\delta}_{nn'}{\bar q}^{mn}
(mn|mn)\,,\,\,(mn|mn)= [m]![n]!\,.
\end{equation}

\section{Physical Subspace of \boldmath{${\cal F}\otimes\bar{\cal F}$}}
\noindent
The study of the physical subspace of the tensor product ${\cal H}\otimes
{\bar {\cal H}}$ of chiral state spaces reduces to the study of its
projection on the finite dimensional space
${\cal F}\otimes\bar{\cal F}$
generated
from the Fock space vacuum by two pairs of creation operators $a^+_\alpha$ and
${\bar a}^\beta_+\,$. We expect that the physical subspace ${\cal F}'
\subset {\cal F}\otimes\bar{\cal F}$
is generated from the $2D$
vacuum (which will be again denoted by
$\vert 0\rangle\,$) by the action of powers of $A^+$
where
\begin{eqnarray}
g(\,(\bar x-x)/2\,,(\bar x+x)/2\,)\,=\,
u_+(x,N){\bar u}^+(\bar x,\bar N)A^++A^-u_-(x,N){\bar u}^-(\bar x,\bar N)+
\nonumber \\
+u_+(x,N)B{\bar u}^-(\bar x,\bar N)-
u_-(x,N+1){^tB}{\bar u}^+(\bar x,\bar N+1)
\end{eqnarray}
with
\begin{equation}
A^\pm=a^\pm_\alpha {\bar a}^\alpha_\pm\quad (\,A^-={^t(A^+)}\,)\,,\quad
B=a^+_\alpha{\bar a}^\alpha_-\,,\quad ^tB=-a^-_\alpha{\bar a}^\alpha_+\,.
\end{equation}
In order to verify the properties of $A^\pm$ and $B$ under transposition
we have to extend the definition of transposition (32) to the algebra of
$U_q^{(2)}$ oscillators setting
\begin{eqnarray}
^ta^+_\alpha ={\cal E}^{\alpha\beta}a^-_\beta\,&,&\quad
^ta^-_\alpha =-{\cal E}^{\alpha\beta}a^+_\beta\,,\, \nonumber \\
^t{\bar a}^{\alpha}_+={\bar a}^\beta_-{\cal E}_{\beta\alpha}\,&,&\quad
^t{\bar a}^{\alpha}_-=-{\bar a}^\beta_+{\cal E}_{\beta\alpha}\,\,.
\end{eqnarray}
The basic CR (12) for the $U_q^{(2)}$ oscillators can be spelled out as follows:
\begin{eqnarray}
a^{\varepsilon}_2a^{\varepsilon}_1=qa^{\varepsilon}_1a^{\varepsilon}_2\,&,&\,
a^-_\alpha a^+_\alpha = a^+_\alpha a^-_\alpha\,,\\
{\bar q}^{1\over 2}a^-_2a^+_1-q^{1\over 2}a^+_1a^-_2={\bar q}^N\,&,&\,
{\bar q}^{1\over 2}a^+_2a^-_1-q^{1\over 2}a^-_1a^+_2=q^N\,,
\end{eqnarray}
and identical relations for ${\bar a}^{\alpha}_\pm\,\,\,
({\bar a}^2_{\varepsilon} {\bar a}^1_{\varepsilon} =
q{\bar a}^1_{\varepsilon} {\bar a}^2_{\varepsilon} \,$ etc.). The following
result is obtained by a direct computation:

\vspace{12pt}
\noindent
{\bf Proposition~4.1:} $\Delta (E)\,,\Delta (F)\,${\it ,
and the bilinear invariant combinations
of $\,a^{\varepsilon}_\alpha\,,{\bar a}^{\alpha}_{\varepsilon}\,$
i.e., $\,A^\pm\,$
and $B\,,\,{^tB}\,$ together with the $q$-exponents $q^{\pm H}\,,
q^{\pm {\bar H}}\,, q^{\pm N}\,, q^{\pm {\bar N}}$ give rise to three
mutually commuting copies of $U_q^{(2)}\,$, their nontrivial CR being}
\begin{equation}
[\Delta (E)\,,\Delta (F)]=\Delta ([H]) = [H+\bar H]\,,\,\,
q^{H+\bar H} \Delta (E) = \Delta (E) q^{H+\bar H +2}\,\, {\rm{etc.}}\,;
\end{equation}
\begin{equation}
[B\,,{^tB}]=[N-\bar N]\,,\,\,q^{N-\bar N} B = B q^{N-\bar N +2}\,\,
{\rm{etc.}}\,;
\end{equation}
\begin{equation}
[A^-\,,A^+ ]=[N+\bar N ]\,,\,\,
q^{N+\bar N}A^\pm = A^\pm q^{N+\bar N \pm 2}\,.
\end{equation}

Taking into account the properties under transposition,
we can say that we obtain the algebra
$U_q(su(2))\otimes U_q(su(2))\otimes U_q(su(1,1))\,$.

\vspace{12pt}
\noindent
We are now prepared to {\it {define the physical subspace}}
${\cal F}'$ of ${\cal F}\otimes{\bar{\cal F}}$ setting
\begin{equation}
\left( (L_\pm )^\alpha_\beta -{\delta}^\alpha_\beta \right){\cal F}' = 0
\end{equation}
(~or $\Delta (E){\cal F}' = 0 = \Delta (F){\cal F}' =
(q^{H+\bar H} -1){\cal F}'\,$);
\begin{equation}
B{\cal F}' = 0 = {^tB}{\cal F}' = (q^{N-\bar N}-1){\cal F}'\,;
\end{equation}
\begin{equation}
(A^-)^{h-1}(A^+)^l{\cal F}' = 0\,,\, l=0,1,\dots ,h-2\,.
\end{equation}

\vspace{12pt}
\noindent
{\bf Proposition~4.2:}
{\it The $U_q(su(2))\otimes U_q(su(2))$ invariant subspace ${\cal F}_I$ 
of the $h^4$ dimensional tensor product
${\cal F}\otimes\bar{\cal F}$
of chiral Fock spaces is $2h-1$ dimensional.
It is generated from the vacuum by the operators
$(A^+)^{(n)}\,,\  n=0,1,\ldots ,2h-2\,$ where}
\begin{equation}
(A^{\pm})^{(n)}=\sum_{k=m}^{n-m}q^{k(n-k)}{(A_1^{\pm})^k\over [k]!}
{(A_2^{\pm})^{n-k}\over [n-k]!}\ ,\ \
A_{\alpha}^{\pm}=a^{\pm}_{\alpha}{\bar a}^{\alpha}_{\pm}\ ,\ \
m=max(0,n-h+1)\ .
\end{equation}
{\it The physical subspace ${\cal F}'$ is $h$ dimensional. It is
spanned by the vectors 
\begin{eqnarray*}
\{\vert n\rangle=
{1\over [n]!}\,(A^+)^n\vert 0\rangle\,,\ \  n=0,1,\ldots ,h-1\}. 
\end{eqnarray*}
The bilinear form induced on ${\cal F}\otimes\bar{\cal F}$ by the
chiral form (36) is
positive semidefinite on ${\cal F}'\,$. Its kernel is one dimensional,
the only zero norm vector in the above basis 
being $\vert h-1\rangle\,$.}

\vspace{12pt}
\noindent
{\bf Proof:} To prove the first statement of the Proposition, we
shall start by stressing that, due to the first Eq.(48), 
$A_2^{\pm}A_1^{\pm}=q^2A_1^{\pm}A_2^{\pm}\ .$ This leads to
$(A^{\pm})^h=0\ ;$ one proves analogously that $B^h=0=(^tB)^h\ .$
However, the (truncated, for $n\ge h\ )$ expression (56) does not
vanish for any $0\le n\le 2h-2$ thus extending unambiguously the
formula for ${(A^{\pm})^n\over [n]!}\ .$ An explicit
calculation shows that the number of the $U_q(su(2))$ scalar
states selected by (53) is $4h-3$. Imposing on them the condition (54) of
the second ("B"-) $U_q(su(2))$ invariance leaves us with a $2h-1$
dimensional subspace spanned by 
\begin{equation}
\{\vert n\rangle=(A^+)^{(n)}\vert 0\rangle\,,\ \ 
n=0,1,\ldots ,2h-2\}\ .
\end{equation} 
To proceed further, we use (52) to prove, for $0\le n\le h-1\ ,$
\begin{equation}
[A^{\mp}\,,(A^{\pm})^{(n)}]=\pm (A^{\pm})^{(n-1)}
[N+\bar N \pm n\mp 1]\ ,
\end{equation}
and the definition (56) to obtain the general 
(for any $0\le n\le 2h-2\ $) relations
\begin{equation}
A^{\pm}\vert n\rangle =[n+1]\vert n\pm 1\rangle\ \ \ 
(\  A^-\vert 0\rangle = 0 =A^+\vert 2h-2\rangle \ ).
\end{equation}
From (59) it follows immediately that $h-1$ among the the basic
vectors (57), those obtained for $n\ge h\ ,$ are eliminated by
condition (55) since
\begin{equation}
(A^-)^{h-1}\vert n\rangle =0\ {\rm for}\  n=0,1,\ldots ,2h-3\ ;\ \ 
(A^-)^{(h-1)}\vert 2h-2\rangle =(-1)^{h-1}\vert h-1\rangle\ .
\end{equation}
This proves the second assertion of the Proposition. 
The last one follows from (58) which implies
\begin{equation}
\langle m\vert n\rangle =[m+1]\,{\delta}_{mn}\ ;\ \ 
m,n=0,1,\ldots ,h-1\ .
\end{equation}

\vspace{12pt}
\noindent
{\bf Remark:}
The operator $(A^-)^{h-1}(A^+)^{h-1}$ is identically zero in ${\cal F}_I\ .$
The vectors (57)
for $n\ge h$ can be obtained alternatively by applying
powers of $A^-$ on the "shadow vacuum vector" corresponding
to the highest possible value $n=2h-2\,$~.

\vspace{12pt}
\noindent
As a corollary, it follows that for the subspace
${\cal H}'\,$ of ${\cal H}\otimes {\bar {\cal H}}\,$
generated by polynomials in $g(t,x)$ acting on the vacuum, the monodromy
invariance condition is satisfied in the weak sense:
\begin{equation}
a^\pm M{\bar M}^{-1}{\bar a}_\pm = a^\pm{\bar a}_\pm q^{\mp (N-\bar N )}
\,\,\,\Rightarrow\,\,\,\{ g(t,x+2\pi )-g(t,x)\}{\cal H}' = 0\,.
\end{equation}
If ${\cal H}''$ is the subspace of ${\cal H}'$ of zero norm vectors then the
{\it {physical subquotient}} is defined, as usual, as the factor space
${\cal H}'/{\cal H}''\,$.

\section{Concluding Remarks}
\noindent
The passage from a chiral WZNW model to the physical space of a diagonal $2D$
theory using the framework of a gauge field theory
has been completed for $G=SU(2)\,$. For the rank $l$ case,
$G=SU(l+1)\,$, one would need $l$ quantum numbers $N_1\,,N_2\,,\ldots\,,
N_l\,$ (instead of the single $N$ of Sec. 4) to label the chiral sectors.
The quantum dimensions of the corresponding representations of
$U_q(su(l+1))$ are given by
\begin{eqnarray*}
{\left( [l]!\right)}^{-1} [N_1]\ldots [N_l][N_1+N_2]\ldots [N_{l-1}+N_l]
\ldots [N_1+\ldots +N_l]\,.
\end{eqnarray*}
Unitarizable representations correspond to non-negative quantum dimensions
i.e., to $(l\le)\linebreak N_1+N_2+\ldots N_l\le h\,$.
The splitting of $u(x)$ into creation and
annihilation parts in the $SU(2)$ case
becomes a decomposition into $l+1$ chiral vertex operators
$u_i (x)\,,\,\,i = 1,2\ldots ,l+1\,$~where~$u_i$ changes
$(N_1,\ldots ,N_l )$ into $(N_1,\ldots ,N_i-1, N_{i+1}+1,\ldots
,N_l )$ ($u_1$ maps it onto $(N_1 +1,N_2,\ldots , N_l)\,,$ and
$u_{l+1}$ -- onto $(N_1,\ldots ,N_l-1)\,$).
One has to introduce $l$ pairs of operators $B_i$
and $^tB_i$ (instead of (46)) in the $2D$ theory, satisfying
\begin{equation}
[B_i\,,{^tB}_j ]= [N_i-{\bar N}_i]\,{\delta}_{ij}\,,\quad
q^{N_i-{\bar N}_i}B_j=B_jq^{N_i-{\bar N}_i+c_{ij}}\,,\quad {\rm{etc.}}
\end{equation}
where $c_{ij}$ is the $su(l+1)$ Cartan matrix.

We hope to return to the problem of extending the results of the present
paper to this more general case in a future publication.


\section* {Acknowledgements}
\noindent
I.T. thanks Anton Alekseev for a stimulating discussion. L.H. thanks
ICTP and INFN, Sezione di Trieste, and I.T. thanks SISSA, Trieste for
hospitality and financial support. The
work of L.H. and I.T. has been supported in part by the Bulgarian
National Foundation for Scientific Research under
contract F--404. I.T. thanks George Pogosyan for
his hospitality in Dubna during the II International Workshop on
Classical and Quantum Integrable Systems where this work was first reported.

\end{document}